\documentclass[letterpaper]{jpconf}

\usepackage{graphicx}
\usepackage{hyperref}
\usepackage{color}
\usepackage{array}
\usepackage{amsmath,amssymb}
\usepackage[normalem]{ulem}

\begin{document}

\title{A new paradigm for petascale Monte Carlo simulation: Replica exchange Wang--Landau sampling}

\author{Ying Wai Li$^{1,2}$, Thomas Vogel$^{1,3}$, Thomas W\"ust$^4$, and David P. Landau$^1$}

\address{$1$ Center for Simulational Physics, The University of
  Georgia, Athens, GA 30602, USA}
\address{$2$ National Center for Computational Sciences, Oak Ridge
  National Laboratory,\break{} Oak Ridge, TN 37831, USA}
\address{$3$ Theoretical Division, Los Alamos National Laboratory,\break{}
  Los Alamos, NM 87545, USA}
\address{$4$ Swiss Federal Research Institute
  WSL, Z\"{u}rcherstrasse 111,\break{} CH-8903 Birmensdorf, Switzerland}

\ead{yingwai.li@mailaps.org}

\begin{abstract}
\noindent
We introduce a generic, parallel Wang--Landau method that is naturally
suited to implementation on massively parallel, petaflop
supercomputers.  The approach introduces a replica-exchange framework 
in which densities of states for overlapping sub-windows in energy
space are determined iteratively by traditional Wang--Landau
sampling. The advantages and general applicability of the method are
demonstrated for several distinct systems that possess discrete or
continuous degrees of freedom, including those with complex free
energy landscapes and topological constraints.
\end{abstract}



\section{Introduction}

One of the great challenges accompanying recent developments in high
performance computing hardware is the question of how to make truly
efficient use of the huge numbers of cores found in today's ``cutting
edge'' machines.  There have been concomitant developments in
methodology in computational statistical physics, one of the most
recent of them being a parallel implementation of the very successful
importance sampling Monte Carlo based approach termed ``Wang--Landau
sampling''~\cite{prl_version}. In the original Wang--Landau sampling,
the \emph{a priori} unknown density of states $g(E)$ of a system is
determined iteratively by performing a random walk in energy space
($E$) and sampling configurations with probability $1/g(E)$ (i.e. with
a ``flat histogram'') \cite{wl_prl,wl_pre,shanho04ajp}.  Many studies
have shown that this procedure is very powerful for studying a number
of diverse problems including those with complex free energy
landscapes because it circumvents the very long time scales
encountered near phase transitions or at low temperatures. The method
also facilitates the direct calculation of thermodynamic quantities,
including the free energy, at any temperature from a \emph{single}
simulation.  (In fact, random walks in the space of more than one
variable allow thermodynamic information to be determined as a
function of more than one thermodynamic field from a single
simulation.)  Wang--Landau sampling is also a generally applicable
Monte Carlo procedure with only a few adjustable parameters, and it
has been applied successfully to a wide range of simulation models,
some with discrete degrees of freedom and some involve degrees of
freedom that are continuous
(see~\cite{rathore02jcp,alder04jsm,taylor09jcp,langfeld12prl} for
examples). While modifications to the initial implementation have been
proposed, e.g.  by optimizing the ``modification factor--flatness
criterion'' scheme~\cite{zhou05pre,zhou06prl,belardinelli07pre}, by
introducing more efficient Monte Carlo trial
moves~\cite{yamaguchi02pre,wu05pre,wuest09prl}, or ``error
correction''~\cite{lee2006}, the underlying simplicity has remained
intact.  Thus, the basic algorithm is an excellent candidate for
parallel implementation.

A few simple attempts at parallelizing Wang--Landau sampling have been
implemented, but these are useful only for a relatively small number
of processors (cores).  One early approach~\cite{wl_pre,shanho04ajp}
subdivided the total energy range into smaller sub-windows, each being
sampled by an independent random walker.  The total simulation time
needed is then limited by the convergence of the slowest walker, but
this can be tuned by an unequal partition of energy
space. Nevertheless, ``perfect'' load balancing is impossible due to
the \emph{a priori} unknown nature of the complex free energy
landscape. Furthermore, the individual energy sub-windows cannot be
reduced arbitrarily in range because some regions of configurational
space would then ultimately become inaccessible.  In another approach to
parallelization, multiple random walkers work simultaneously on the
\emph{same} density of states and histogram. Distributed memory
(MPI~\cite{khan05jcp}), shared memory (OpenMP~\cite{zhan},
multi-thread~\cite{wang}), and GPU~\cite{yin12cpc} variants of this
idea have been proposed; shared memory implementations have the
advantage of not requiring periodic synchronization among the walkers
and even allow for ``data race'' when updating $g(E)$ without
noticeable loss in accuracy~\cite{zhan}.  Although this second
approach would appear to avoid the problems of the first approach, a
recent, massively parallel implementation~\cite{yin12cpc} found that
correlations among the walkers could lead to a systematically
underestimation of $g(E)$ in ``difficult to access'', low energy
regions. The addition of a phenomenological bias to the modification
factor alleviated this difficulty, the effective round-trip times of
the individual walkers, however, are not improved through the use of
this "ad hoc" method and the validity of the approach cannot be confirmed.

\section{The new, parallel (replica exchange Wang--Landau) algorithm}

Our new approach~\cite{prl_version,tv_follow_2013} is a
\emph{generic}, parallel Wang--Landau scheme which combines the
benefits of the original Wang--Landau (WL) sampling scheme with those
of replica-exchange Monte Carlo \cite{partemp1,partemp2,partemp3}.
Much of the success of the Wang--Landau algorithm has resulted from
the combination of its simplicity and robustness, and our goal was to
retain these qualities in the parallel algorithm.  We begin by
splitting up the total energy range into smaller, overlapping
sub-windows.  Sampling then proceeds in each energy sub-window by
multiple, \textit{independent} Wang--Landau walkers, each of which has
its own instantaneous density of states and histogram.  The key to
this approach is that configurational, or replica, exchanges are
allowed between walkers in overlapping energy sub-windows
during the simulation.  Consequently, each replica can travel back and
forth over the entire energy space many times during a single
simulation. The replica exchange move does not bias the
overall procedure; thus the approach is applicable with any valid
trial update or Wang--Landau convergence criterion (e.g., the ``$1/t$
algorithm''~\cite{belardinelli07pre}).  Furthermore, the parallel
algorithm does not impose any intrinsic limitation to the number of
random walkers and the number of energy sub-windows that can grow with
the size of the system. Therefore, the computational method should
scale straightforwardly to many thousands of cores.

The standard Wang--Landau algorithm~\cite{wl_prl,wl_pre} estimates the
density of states, $g(E)$, using a single random walker in an energy
range $\left[ E_{\mathrm{min}}, E_{\mathrm{max}}\right]$. During the
simulation, trial moves are accepted with a probability $P =
\textrm{min} \left[1, g(E_\mathrm{old}) / g(E_\mathrm{new})\right]$,
where $E_\mathrm{old}$ ($E_\mathrm{new}$) is the energy of the
original (proposed) configuration. The estimation of $g(E)$ is
continuously adjusted and improved using a modification factor $f$
(i.e. $g(E) \to f \times g(E)$) which starts with $f_0 > 1$ and
progressively approaches unity as the simulation proceeds. A
histogram, $H(E)$, keeps track of the number of visits to each energy
$E$ during a given iteration.  When $H(E)$ is sufficiently ``flat'',
the next iteration begins with $H(E)$ reset to zero but keeping the
estimate of $g(E)$ from the previous iteration, and $f$ reduced by
some predefined rule (e.g. $f \rightarrow \sqrt f$). The simulation
ends when $f$ reaches a sufficiently small value $f_\mathrm{final}$ at
which point the accuracy of $g(E)$ is proportional to $\sqrt
f_\mathrm{final}$ for sufficiently flat $H(E)$~\cite{zhou05pre}.

\begin{figure}
\centering
  \includegraphics[width=0.55\columnwidth,clip]{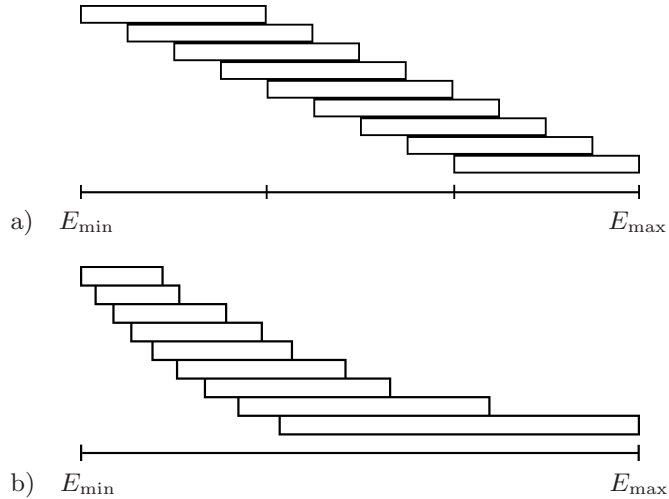}
  \caption{\label{fig:1}%
    a) Subdivision of the global energy range into nine equal-size
    intervals with overlap $o = 75\%$. b) Run-time balanced subdivision
    with overlap to the higher energy interval $o\geq75\%$. Multiple
    Wang--Landau walkers can be employed within each interval.}
\end{figure}

In our parallel Wang--Landau framework, the global energy range is
first split into $h$ over\-lapping sub-windows, and $m$ random walkers
are used to sample each sub-window.  The extent of the overlap $o$
should strike a balance between fast convergence of $g(E)$ and a
reasonable exchange acceptance rate.  While an overlap of
$o\approx75\%$ works well, excellent results can be obtained with
other values~\cite{tv_follow_2013}.  (Different types of partitioning
of the global energy into sub-windows may be used; but, in all cases,
configurational exchange cannot occur without overlap of adjacent
sub-windows, see Fig.~\ref{fig:1} for examples.)  Within an energy
sub-window, each random walker performs standard Wang--Landau sampling.
After a predetermined number of Monte Carlo steps, a ``replica
exchange'' is proposed between two random walkers, $i$ and $j$, where
walker $i$ chooses swap partner $j$ from a neighboring sub-window at
random. For the sake of simplicity and generality in our discussion,
we do not sort out walkers which are currently not in the overlap
region before drawing the pairs. Let $X$ and $Y$ be the
configurations that the random walkers $i$ and $j$ are ``carrying''
before the exchange; $E(X)$ and $E(Y)$ be their energies,
respectively.  From the detailed balance condition the acceptance
probability $P_{\mathrm{acc}}$ for the exchange of configurations $X$
and $Y$ between walkers $i$ and $j$ is:
\begin{equation}
  P_{\mathrm{acc}}=\min\left[1,\frac{g_i(E(X))}{g_i(E(Y))}\frac{g_j(E(Y))}{g_j(E(X))}\right]\,
\end{equation}
where $g_i(E(X))$ is the instantaneous estimator for the density of
states of walker $i$ at energy $E(X)$, cf.~\cite{nogawa11pre}.  Note
that if either of the random walkers has an energy that lies outside
the range of the sub-windows of the other, the replica exchange cannot
take place. In that case, we just disregard the exchange and carry out
another number of MC steps until the next replica exchange is
proposed. This combination of sampling trials defines the new parallel
algorithm.

An important new feature of our formalism is that each random walker
has its own $g(E)$ and $H(E)$ which are updated independently.  Also,
since every walker has to satisfy the Wang--Landau flatness criterion
\emph{independently} at each iteration, the systematic errors found
in~\cite{yin12cpc} are avoided. When all random walkers within an
energy sub-window have attained flat histograms, their estimates for
$g(E)$ are averaged and then redistributed among themselves before
simultaneously proceeding to the next iteration. This procedure
reduces the error during the simulation with
${\sqrt{m}}$~\cite{tv_follow_2013}, i.e. as for uncorrelated WL
simulations. Furthermore, increasing $m$ can improve the convergence
of the Wang--Landau sampling by reducing the risk of statistical
outliers in $g(E)$ resulting in slowing down subsequent
iterations. Alternatively, it allows us in principle, to use a
weaker flatness criterion~\cite{tv_follow_2013}, which is in the
spirit of a concurrently proposed idea of merging histograms in
multicanonical simulations~\cite{zierenberg13cpc}.

The parallel simulation ends when the modification factors in all the
energy intervals have reached $f_\mathrm{final}$ and the $h\times m$
pieces of $g(E)$ fragments with overlapping energy sub-windows are
then used to construct a single $g(E)$ over the entire energy
range.

\section{Data analysis and production run}

\begin{figure}
\centering
  \includegraphics[width=0.55\columnwidth,clip]{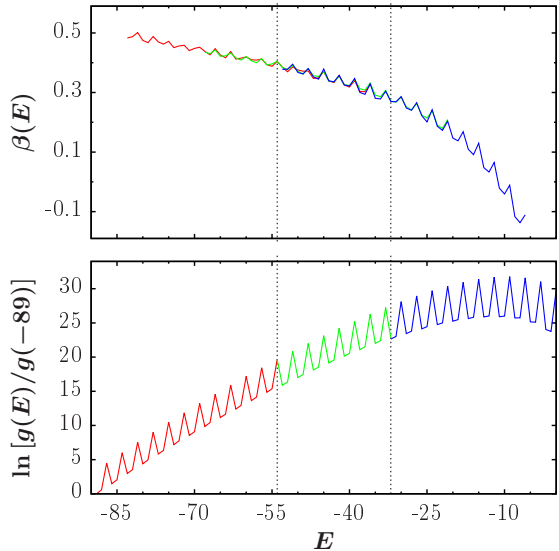}
  \caption{\label{fig:2}%
    (color) Joining pieces of $g(E)$ for the HP 67mer adsorption on a
    weak attractive surface (Complete data shown in
    Fig.~\ref{fig:4}\,a). (Top) First derivatives of raw DOS pieces
    from the three highest-energy windows. We applied a 5-point
    stencil with step-width $\Delta E = 3$.  Derivatives coincide best
    at $E = -54$ (red vs. green curve) and at $E=-32$ (green vs. blue
    curve; marked by vertical dotted lines). (Bottom) Pieces are
    connected at these points to obtain overall DOS.}
\end{figure}

To connect two pieces of consecutive, overlapping density of states
fragments, say $g_i(E)$ and $g_j(E)$, we first
calculate the inverse microcanonical temperatures, $\beta_i(E)$ and
$\beta_j(E)$ by: $\beta(E) =\textrm{d}\log[g(E)]/\textrm{d}E$. The
joining point, $E_{\mathrm{join}}$, is determined as the point where
$\Delta \beta = |\beta_i(E) - \beta_j(E)|$ vanishes or is the
smallest. This procedure avoids discontinuities in the
derivatives of the final density of states, which can lead to
artificial peaks in derived functions like the heat capacity, for
example. Next, $g_j(E)$ has to be rescaled using the value
$g_i(E_{\mathrm{join}})$ as the reference point to yield a correct
ratio for different energy levels: $g_j(E) \rightarrow g_j(E)
(g_i(E_{\mathrm{join}})/g_j(E_{\mathrm{join}}))$. Finally, a
joined relative density of states can be obtained by:
\begin{equation}
\label{eq:joinDOS}
g(E) = \left\{
\begin{array}{ll}
  g_i(E) & \textrm{if $E < E_{\mathrm{join}}$} \\
  g_j(E) & \textrm{if $E \geq E_{\mathrm{join}}$}\,.
\end{array}
\right.
\end{equation}
Here, we have assumed $j > i$ for Eq.~(\ref{eq:joinDOS}) but this does
not need to be the case for the previous rescaling procedure to work
properly. See Fig.~\ref{fig:2} for an illustration of the procedure
described so far. Since there are $m$ pieces of density of states in
each energy window, it is possible to compute the statistical errors
by standard resampling techniques, e.g. jackknife or bootstrap
methods, which can be found in standard textbooks. A more
comprehensive discussion on this topic as applied to our data analysis
scheme is presented in Ref. \cite{tv_follow_2013}.

Very often one is not only interested in obtaining a one-dimensional
density of states, but a two-dimensional (or even higher dimensional)
density of states, $g(E, Q)$, where $Q$ is a~physical quantity. Such
joint densities of states can be particularly useful for the
calculation of thermodynamic properties of order parameters, for
instance. One way of doing this is to perform a two-dimensional
Wang--Landau sampling as proposed in the original
treatment~\cite{wl_prl,wl_pre,shanho04ajp}; a more efficient way is to
carry out a two-dimensional multicanonical production run using
$1/g(E)$, obtained from a one-dimensional Wang--Landau sampling, as
the simulation weight and construct $g(E, Q)$ from measurements of $E$
and $Q$ (see, for example, Ref.~\cite{Li2012} for details).

In any case, one is confronted with the necessity of constructing the
entire two-dimensional, joint density of states from fragments of
$g(E, Q)$. We found that a direct generalization from the
aforementioned one-dimensional scheme with a slight modification is
able to yield satisfactory outcomes\,\footnote{The construction of a
  higher-dimensional density of states, in principle, could follow the
  same fundamental scheme as far as the calculation of the normal
  vector (Eq.~\ref{eq:normvec}) is properly generalized to higher
  dimensions.}:  let $g_i(E, Q)$ and $g_j(E, Q)$ be the two pieces of
2D density of states to be merged. We first calculate the normal
vectors of the surfaces $S(E, Q)=\log\,g(E, Q)$:
\begin{equation}
  \hat{N} = \left(
  \begin{array}{c}
    \partial S(E, Q)/\partial E \\[1mm]
    \partial S(E, Q)/\partial Q \\[1mm]
    -1 
  \end{array}
  \right)\,,\label{eq:normvec}
\end{equation}
from which the unit normal vectors $\hat{n}$ are calculated at all
points $(E, Q)$. The joining position can then be determined as the
point where these vectors best coincide, i.e., where $\hat{n_i} \cdot
\hat{n_j}$ is maximal. One of the $g(E, Q)$ fragments is again
rescaled in the same way as in the 1D case, and finally a joined
relative density of states is determined by:
\begin{equation}
\label{eq:join2DDOS}
g(E, Q) = \left\{
\begin{array}{cl}
  g_i(E, Q) & \textrm{if $(E, Q)$ is present only in window $i$} \\
  g_j(E, Q) & \textrm{if $(E, Q)$ is present only in window $j$}\\
  (g_i(E, Q) + g_j(E, Q))/2  & \textrm{if $(E, Q)$ is present in
   both windows} \,.
\end{array}
\right.
\end{equation}
This procedure is used for the analysis of our example models in
Sect.~\ref{sec:vistas}.

\section{Performance measures and scaling}

In order to assess the generality and performance of this parallel
Wang--Landau scheme, we applied it to multiple, fundamentally different
models.\footnote{We emphasize again, however, that there is nothing in
  our framework that restricts its use to these models!} As for most
new computational schemes in statistical physics, we first tested our
framework on the ``fruit fly'' of statistical physics, the Ising
model. We simulated the 2D Ising model with system sizes as large as
$256^2$ using more than 2000 cores. While single-walker Wang--Landau
sampling typically takes more than a week to converge, the parallel
scheme easily finishes within a few hours, with the same accuracy as
the serial scheme. The deviations from the exact results are always of
the same order as the statistical errors, which are $<0.01\%$ in the
peak region of the density of states.

\begin{figure}
\centering
  \includegraphics[width=.45\textwidth]{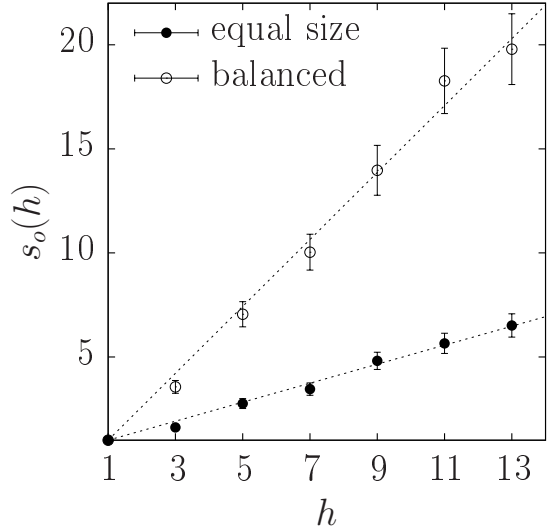}
  \caption{\label{fig:5} a) Dependence of the speed-up $s_o(h,m)$ on
    the number of energy windows and overlap $o = 75\%$ (filled
    symbols, cf.  Fig.~\ref{fig:1}\,a) and using a run-time balanced
    energy splitting (open symbols, cf.  Fig.~\ref{fig:1}\,b). Here,
    the calculation of the speed-up is based on the MC steps (MCS)
    needed to complete the first WL iteration. Measurements are
    performed using the lipid bilayer system, see text for details.  }
\end{figure}
To generally quantify the efficiency of the parallel WL scheme, we
define the speed-up, $s_o(h,m)$, as the number of Monte
Carlo steps taken by the slowest parallel WL walker
($N_o^{\mathrm{parallel}}(h,m)$), as compared to that taken by a
single walker ($N^{\mathrm{single}}$):
\begin{equation}
\label{eq:speed}
s_o(h,m) =\frac{N^{\mathrm{single}}}{N_o^{\mathrm{parallel}}(h,m)}.
\end{equation}
We measure this number for $h\lesssim20$ in simulations of the lipid
system to be introduced below. As shown in Fig.~\ref{fig:5}, we find
strong scaling, i.e.  the speed-up scales \textit{linearly} with $h$
for a fixed number $m$ for both energy splittings as shown in Fig.
\ref{fig:1}.  While the equal-size energy range splitting
(Fig.~\ref{fig:1}\,a) is the most basic approach, the run-time
balanced energy splitting (Fig.~\ref{fig:1}\,b) is chosen such that
walkers in different energy sub-windows complete the first WL
iteration after the same number of MC sweeps (within statistical
fluctuations). As the growth behavior of WL histograms is in principle
known \cite{zhou05pre}, such an energy splitting can be estimated by
analyzing the first-iteration histogram from a short pre-run with
equal-size energy intervals.\footnote{Of course, the energy-window
  setup can be further adapted as the simulation proceeds.} We have
shown in Ref.~\cite{prl_version} that our method can also achieve weak
scaling properties. With the capability of achieving both strong and
weak scaling (i.e., by increasing the number of computing cores one
can get results faster for the same system and/or simulate larger
systems in the same period of time, respectively), our formalism
becomes a promising candidate for large-scale applications.

\section{Opening new vistas}
\label{sec:vistas}

To show how this parallel framework can allow us to examine previously
unapproachable problems, we also applied the method to two very
distinct and particularly challenging molecular systems: a
coarse-grained continuum model for the self assembly of amphiphilic
molecules (lipids) in explicit water and a discrete model for the
surface adsorption of lattice proteins. In the first model,
amphiphilic molecules, each of which is composed of a
polar (P) head and two hydrophobic (H) tail monomers (P--H--H), are
surrounded by solvent particles (W). The interactions between H and W
molecules, as well as those between H and P molecules, are purely
repulsive. All other interactions between non-bonded particles are of
Lennard-Jones type; bonded molecules are connected by a FENE
potential, see Refs.~\cite{getz,fujiwara} for similar models. The
second model uses the hydrophobic-polar
(HP) lattice model~\cite{Dill1985} for protein surface adsorption.
Here a protein is represented by a self-avoiding walk consisting of H
and P monomers placed on a simple cubic lattice with an attractive
substrate. For recent simulational results on this model and
computational details see~\cite{Li2012,Li2011}.

\begin{figure}
\centering
  \includegraphics[width=0.64\textwidth,clip]{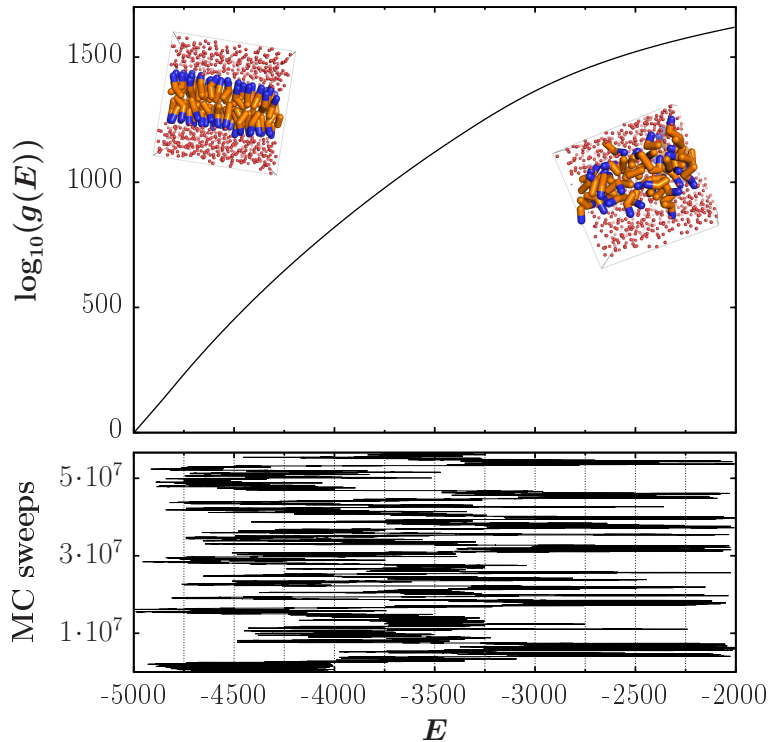}
  \caption{\label{fig:3}%
    (color; Top) Logarithm of the density of states of the amphiphilic system
    containing 75 lipid molecules and a total of 1000 particles
    obtained by our parallel Wang--Landau scheme with the setup shown
    in Fig.~\ref{fig:1}\,a. The pictures show a conformation
    where lipid molecules assemble and form a single cluster
    ($E\approx-2100$) and a low-energy bilayer configuration
    ($E\approx-4800$). (Bottom) Path of one single replica (out of 27
    contributing to the above density of states) through energy space.
    Replica exchanges between walkers are proposed every $10^4$ sweeps
    (data also shown with that resolution), with acceptance rates
    between 30 and 55\,\%. Grid lines correspond to the borders of the
    individual energy windows.}
\end{figure}
Both models pose technical challenges due to high energy and$/$or
configurational barriers. Simulations of either setup is impossible
for all practical purposes using the traditional, single walker
Wang--Landau method due to unreasonable resource demands. (Standard
Metropolis sampling would be \emph{many} orders of magnitude too
slow.) The first model (the amphiphilic system) consists of 75 lipid
molecules (each composed of three particles) and 775 solution
particles with a continuous energy domain.  The density of states
$g(E)$ over an energy range covering the lipid bilayer formation spans
more than 1600 orders of magnitude, which makes precise low
temperature properties extremely difficult to estimate. Only with the
parallel algorithm, it is possible to clearly observe and study the
lipid bilayer formation and different bilayer phases in this system.
See Fig.~\ref{fig:3} for example data and
Refs.~\cite{vanderbilt_jcp,brazil_proceeding} for all simulation
details.

\begin{figure}
\centering
  \includegraphics[width=0.9\textwidth,clip]{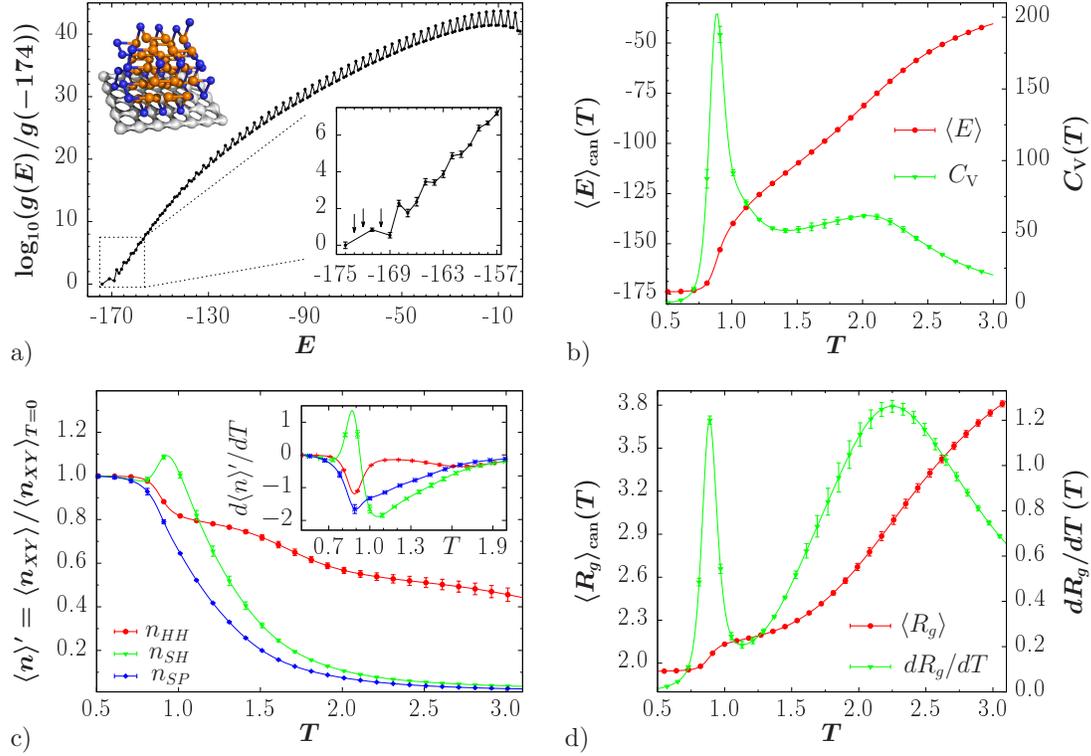}
  \caption{\label{fig:4}%
    (color; a) Density of states of the lattice HP 67mer, where only
    H-monomers are attracted by the substrate. The H--H interaction is
    three times stronger than the surface attraction
    ($\varepsilon_{HH} = 3$, $\varepsilon_{SH} = 1$ and
    $\varepsilon_{SP} = 0$), leading to the unusual sawtooth like
    shape. The inset shows the error bars on the enlarged low-energy
    data. Note the two energy gaps, i.e. no conformations exist with
    $E=-173$, $-172$, and $-170$ (see arrows). The picture shows an
    adsorbed HP protein with energy $E=-174$. (b) The corresponding
    canonical mean energy and heat capacity; (c) number of single H--H
    interactions and surface contacts; and (d) radius of gyration.}
\end{figure}

The second model (the HP lattice protein) under consideration
consists of 67 monomers \cite{yue95pnas} interacting with a weakly
attractive surface, for which the total energy of the system can be
calculated as:
\begin{equation}
  \label{eq:HPenergy}
  E = -(n_{HH} \varepsilon_{HH} + n_{SH} \varepsilon_{SH} + n_{SP} \varepsilon_{SP}),
\end{equation} 
where $n_{HH}$, $n_{SH}$ and $n_{SP}$ are the number of hydrophobic
interactions, surface-H interactions, and surface-P interactions,
respectively; $\varepsilon_{HH}$, $\varepsilon_{SH}$ and
$\varepsilon_{SP}$ are the corresponding energy scales. The discrete
energy levels and the unequal interaction strengths between the H
monomers and with the surface result in an unusual, sawtooth-like
density of states, as shown in Fig.~\ref{fig:4}\,a. This, combined
with other obstacles such as the first-order like structural
transitions in the system and the peculiar form of the density of
states near the ground state, makes convergence for the entire energy
range extremely time-consuming using only a single walker.  Simulating
smaller energy sub-windows individually as proposed by earlier studies
was also not successful for this kind of system, due to the fact that
too-small energy window would destroy ergodicity in the simulation.
Consequently, some regions in the configurational space are
unreachable, resulting in an incompletely simulated density of
states.\footnote{This is not to be confused with the energy gaps shown
  in Fig.~\ref{fig:4}\,a, as those are \textit{real} missing energy
  levels for this system.}

However, we successfully simulated this model by our parallel
Wang--Landau algorithm within a reasonable time frame. The density of
states over the entire energy range and the thermodynamic properties
of a few structural properties are shown in Fig. \ref{fig:4}. From
$g(E)$, the average energy $\langle E \rangle$ and the heat capacity
$C_V$, both shown in Fig. \ref{fig:4}\,b, can be calculated:
\begin{align}
\label{energy}\left\langle E \right\rangle &= Z^{-1}\sum_{E}{E g(E) e^{-E/k_B T}}\,,\\
C_V &= \frac{\left\langle E^2 \right\rangle - \left\langle E \right\rangle^2}{k_B T^2}\,,
\label{Cv}
\end{align}
where $Z = \sum_{E}{g(E) e^{-E/k_B T}}$ is the partition function at
temperature $T$ with Boltzmann factor~$k_B$.

To identify precisely which transitions are responsible for the peak
at $T \approx 0.9$ and the shoulder across $T \approx 1.0$--$2.0$ in
$C_V$, we compare it with the thermodynamics of the number of contacts
between different particle species, i.e., the contributions to the
total energy given in Eq.~(\ref{eq:HPenergy}), and the radius of
gyration, $R_g^2 = \frac{1}{N}\sum_{i=1}^N{(\vec{r}_i -
  \vec{r}_{cm})^2}$, where $\vec{r}_{cm}$ is the center of mass of the
configuration, $\vec{r}_i$ is the position of monomer $i$, and $N$ is
the chain length. These results are shown in Fig. \ref{fig:4}\,c and~d
and were obtained through the joint density of states estimation as
discussed earlier, where mean values are calculated via:
\begin{equation}
  \label{eq:meanval}
  \langle Q\rangle=Z_Q^{-1}\sum_{E,Q}Q\,g(E,Q)\,e^{-E/k_BT}\,,
\end{equation}
using the corresponding partition sum $Z_Q$.

From the derivative of the number of H--H contacts, $d \left\langle
  n_{HH} \right\rangle / dT$, we can see that the hydrophobic core
formation occurs mainly at $T \approx 0.9$ and mildly at $T \approx
1.7$. By looking at the number of surface--H contacts, $d\left\langle
  n_{SH} \right\rangle / dT$, it is clear that adsorption is initiated
by the attraction of the hydrophobic surface at $T \approx 1.1$. The
polar monomers, although not attracted by the surface, are dragged to
come in contact with the surface besides forming a polar shell of the
protein. This causes a negative peak in $d\left\langle n_{SP}
\right\rangle / dT$ at a lower temperature of $T \approx 0.9$. But
recall that at the same temperature, the transition to ground state
(hydrophobic core formation) is also taking place, which draws some H
monomers from the surface to the core, resulting in a negative trough
(visually, a peak) in $d\left\langle n_{SH} \right\rangle / dT$.

The radius of gyration and its derivative, shown in
Fig. \ref{fig:4}\,d, give us extra information about the structural
changes of the system. The peak at $T \approx 2.2$ in $d\langle R_g
\rangle / dT$ signals the $\theta$-transition\footnote{For a review of the
    $\theta$-transition in lattice-polymer models, see, for example,
    Ref.~\cite{vogel07pre}.}, where an extended coil collapses into a globular
structure. This confirms the above observation on $d\left\langle
  n_{HH} \right\rangle / dT$, as it is necessary to first (i.e., at a
higher temperature) bring the monomers closer together before the mild
hydrophobic core formation can take place at $T \approx 1.7$. Another
peak found at $T \approx 0.9$ is a more obvious
reinforcement of our previous observation. During the transition to
the ground state, there is a competition between the major hydrophobic
core formation (which tends to decrease $R_g$ significantly) and
adsorption (which tends to increase $R_g$ slightly), causing a large
fluctuation in $R_g$. Since both the hydrophobic core formation and
adsorption take place at nearly the same temperature, a single,
pronounced peak is observed in $C_V$ eventually. This is a typical
Category~III transition according to the scheme defined in Ref.~\cite{Li2012}.

Using the lipid system as an example and considering a much smaller
global energy range accessible for single-walker simulations, we
measure the speed-up defined by Eq.~(\ref{eq:speed}). The slope of the
speed-up in completing the first iteration is $\approx 0.5$ for the
equal size energy splitting and $\approx 1.6$ for the run-time
balanced energy splitting (cf.  Fig.~\ref{fig:5}), which is
particularly remarkable as this indicates that the speed-up is
\emph{larger} than the number of processors used.  For the HP protein,
even a basic set-up of equal-size energy splitting with only a single
walker per energy interval yields a speed-up of $s_{o = 75\%}(h = 9)
\approx 20$ compared to single walker Wang--Landau simulations. Again,
we get a speed-up larger than the number of
processors.\footnote{Note that we consider the effect of our
    parallel scheme as a \emph{whole}, compared to the original,
    single-walker Wang--Landau method. That is, there are multiple
    factors contributing to the speed-up as defined here. In a purely
    technical definition, where algorithmic and methodological
    influences are absent, speed-ups larger than the number of used
    processors are obviously unreachable.} It is conceivable that the
speed-up factor is ``mysteriously'' larger than the number of
processors since the sum of entries needed to create flat histograms
in all small energy windows can well be, and in the case of the
balanced energy range splitting indeed is (data not shown), smaller
than the number needed for a flat histogram on the whole energy space.
Furthermore, our parallel scheme combines the advantages of two levels
of parallelism which both contribute to the acceleration: first, as
just mentioned, each walker only needs to attain a flat histogram in a
smaller energy window; second, the replica exchange process can
revitalize walkers from trapped states, thus shortening the time spent
on redundant sampling of rare events.  It also avoids an erroneous
bias in $g(E)$ due to potential ergodicity breaking, as replicas can
access the \textit{entire} conformational space by walking through all
energy windows (a typical time-series of a replica performing round
trips in the full energy range of the lipid system is shown at the
bottom in Fig.~\ref{fig:3}). A more detailed analysis and discussion
can be found in Ref.~\cite{tv_follow_2013}.

\section{Summary and Outlook}

In summary, we introduced a generic, parallel framework for
generalized ensemble Wang--Landau simulations making use of energy
range splitting, replica exchange, and multiple random walkers. The
method is simple, general, and leads to significant advantages over
the traditional, serial algorithm. In our complete formulation, we
consider multiple Wang--Landau walkers in independent parallelization
directions and show that both strong and weak scaling can be
achieved.  With the ability to produce highly accurate results, and
proven scalability up to $\approx 2000$ cores without introducing any
bias, we have demonstrated that our parallel scheme has the potential
for extremely large scale parallel Monte Carlo simulations. Since the
framework is complementary to other technical parallelization
strategies, it is further extendible in a straightforward manner. This
facilitates efficient simulations of larger and more complex systems,
and thus provides a basis for many applications on petaflop machines
and beyond.

\ack {We thank M. Eisenbach and E.A. Zubova for constructive
  discussions. This work was supported by the National Science
  Foundation under Grants DMR-0810223 and OCI-0904685. Y.W. Li was
  partly sponsored by the Office of Advanced Scientific Computing
  Research; U.S. Department of Energy. Part of the work was performed
  at the Oak Ridge Leadership Computing Facility at ORNL, which is
  managed by UT-Battelle, LLC under Contract No. De-AC05-00OR22725.
  Supercomputer time was also provided by TACC under XSEDE grant
  PHY130009. Assigned: LA-UR-13-28353.}

\end{document}